# De Novo Assembly of Uca minax Transcriptome from Next Generation Sequencing

Hanin Omar[1], Casey A. Cole[1], Arjang Fahim[1], Giuliana Gusmaroli[2], Stephen Borgianini[2], and Homayoun Valafar[1*]

[1]Department of Computer Science and Engineering, University of South Carolina, Columbia, SC 29208, USA

[2]Department of Natural Sciences, University of South Carolina, Beaufort, SC 29901, USA

* Corresponding Author Email: homayoun@cec.sc.edu Phone: 1 803 777 2404 Fax: 1 803 777 3767

Mailing Address: Swearingen Engineering Center, Department of Computer Science and Engineering, University of South Carolina, Columbia, SC 29208, USA

**Abstract** – *High-throughput cDNA sequencing (RNA-seq) is a very powerful technique to quantify gene expression in an unbiased way. The Crustacean family is among the groups of organisms sparsely represented in current genomic databases. Here we present transcriptome data from Uca minax (red-jointed fiddler crab) as an opportunity to extend our knowledge. Next generation sequencing was performed on six tissue samples from Uca minax using the Illumina HiSeq system. Six Transcriptome libraries were created using Trinity; a free, open-source software tool for de novo transcriptome assembly of high-throughput mRNA sequencing (RNA-seq) data with the absence of a reference genome. In addition, several tools that aid in management of data were used, such as RSEM, Bowtie, Blast, and IGV; a tool for visualizing RNA-seq analysis results. Fast quality control (FastQC) analysis of the raw sequenced files revealed that both adapter and PCR primer sequences were prevalently present, which may require a preprocessing step.*

## 1 Introduction

The advent of Structural Genomics era, marked by completion of the Human Genome project in 2003[1], [2] has introduced many exciting avenues of research in molecular biology and understanding of diseases. Assisted by revolutionizing technologies such as Polymerase Chain Reaction (PCR)[3], and advances is instrumentation[4][5], genome sequencing continues to become a more routine operation compared to other areas of molecular sciences such as protein structure determination[6][7], [8]. While the initial cost of human genome sequencing consisted of $10^9$, the new technologies afford a reduced cost of $10^3$ for sequencing of the same genome. In addition, from the temporal perspective, the initial human genome project required 10 years of data acquisition and three years of data analysis. In contrast, the use of Next Generation Sequencing (NGS)[9], [10] technology has reduced this time requirement to less than a week for combined data acquisition and analysis.

These significant advances in financial and temporal cost of gene sequencing is rooted in development of massively parallel instance of sequencing shorter sequences (100-10,000 bp). While the new parallel approaches increase the overall throughput by several orders of magnitude, they impose the challenge of gene assembly from short reads. Therefore, in recent years one visible area of research and development is focused on evaluation and development of new assembly techniques.

Computational approaches to assembly of NGS sequences can be place in two distinct categories: "Mapping first", and De novo assembly. The former approach relies on existence of a sequenced genome of an organism closely related to the organism under examination. In this approach the existing genome is used as a template to map, following the assembly of the NGS reads. The former approach aims to assemble the short sequences first without any other a-priori knowledge. Programs such as AbySS[11], SOAPdenovo[12], Oases[13] or Trinity[14] can be cited as De novo approaches to sequence assembly. Programs such as Scripture[15] or Cufflinks[16] can be cited as this category of tools.

In this research we utilized the Trinity software because of its availability, popularity, performance, and hardware requirement in order to assemble the transcriptome of the organism Uca minax. Our selection of Uca minax is based on a number of its unique biological properties including adaptability to a broad spectrum of salinity in its environment. Furthermore, there has been a very poor sampling for the genome of crustaceans and therefore assembled transcriptome of Uca minax will help to close this gap in genetic information. In this report we present the assembly results of transcriptome from six tissues of Uca minax using the software package Trinity.

## 2 Materials and Methods

### 2.1 Next generation sequencing data

Messenger RNA (mRNA) was extracted from the following six tissues of Uca minax: anterior gills, posterior gills, gonads (male and female), eye stalk and muscle, first and second Zoea stage, and third Zoea stage using oligodT primer. The mRNA samples were then fragmented and reverse transcribed using random primers. Next, they are double stranded and ligated with adapters according to the Illumina (http://www.illumina.com) prep library protocol. This process yielded six standard cDNA libraries (one for each tissue) for sequence analysis via Illumina HiSeq™ 2000 (commercial service) at the Genomics LAB of the David H. Murdock Research Institute (DHMRI) in North Carolina (www.http://dhmri.org/). These libraries were then combined together to generate a single library pool for the sequencing exercise. This pool was loaded into one lane of the flow cell. Clusters were generated for a paired end read flow cell and a 100 Bp strand specific paired end read sequencing was performed. This process yielded a total of twelve fastq files; two for each library. Table 1 displays the number of raw reads and corresponding size of each sequenced tissue.

*Table 1: Uca minax raw reads sequenced by NGS*

| Tissue Name | File type | Raw Reads count | Size (GB) |
|---|---|---|---|
| Anterior Gills | R1 | 7360115 | 1.9 |
|  | R2 | 7360115 | 1.9 |
| Posterior Gills | R1 | 10870260 | 2.8 |
|  | R2 | 10870260 | 2.8 |
| Female + Male Gonads | R1 | 8308628 | 2.1 |
|  | R2 | 8308628 | 2.1 |
| Eye Stalk + Muscle | R1 | 9761180 | 2.5 |
|  | R2 | 9761180 | 2.5 |
| 1st + 2st Zoea Stage | R1 | 8481444 | 2.2 |
|  | R2 | 8481444 | 2.2 |
| 3rd Zoea Stage | R1 | 9973433 | 2.6 |
|  | R2 | 9973433 | 2.6 |

Image processing and base calling steps were performed at DHMRI to generate the following summary report of sequencing data quality control:

1. Q score: > 80% of bases had a quality score (Q) > 30. The quality or $Q_{score}$ is defined in Eq. (1) and measures the probability that a base is called incorrectly. A Q score of 30 reflects the probability of an incorrect base call of 1 in 1000 for an inferred base call accuracy rate of 99.9%.

$$Q = -10 \log_{10}(e) \qquad \text{Eq. (1)}$$

2. Data throughput: The data throughput quality control (QC) threshold was set to be larger than 100 million reads/lane.

## 2.2 De novo sequence assembly

The Trinity[14][17] software package (release 2013-2-25) optimized with k-mer[18] length of 25 for performing de novo assembly on the raw reads was used in this work. The computational work was performed on a plank cluster with 864 cores, and each node populated with 24 GB of RAM memory. The operating system of this computational facility was CentOS (https://www.centos.org/). The command-line arguments used with Trinity are shown in Dialogue 1. Each of the parameters is briefly described in Table 2.

```
---left    "compatible_path_extension_for_reverse_reads"
--right   "compatible_path_extension_for _forward_reads"
--seqType fq --SS_lib_type RF  --JM 20 --CPU 12 –Output
"compatible_path_extension_for _output_folder".
```

*Dialogue 1. Command line arguments used during assembly session with Trinity.*

*Table 2. Arguments used with Trinity and a brief description of each.*

| Argument | Brief Description |
|---|---|
| ---left | Input file name for left reads |
| --right | Input file name for right reads |
| --seqType | Type of input files; fastq or fasta |
| --SS_lib_type | Define the left and right files read orientation, RF or FR |
| --JM | The memory assigned for the kmer dictionary in GB |
| --CPU | Number of CPU assigned for Trinity to use |
| -Output | Name of the output folder that contains the assembled transcriptome file |

Due to the ambiguity of the protocol used in the preparation of the strand-specific libraries for sequencing, we ran Trinity twice for each tissue. In the first run the –SS_lib_type parameter was set to FR, and on the second run it was set to RF. To explore the performance of Trinity on these datasets we varied the number of cores being utilized by the process and recorded the run time. Table 3 shows the effect of this variation of cores assigned to Trinity on the overall running time performance.

*Table 3: Trinity run time analysis*

| Tissue name | Numbers of cores | Total running time (hours) |
|---|---|---|
| Anterior gills | 6 | < 7 |
|  | 12 | < 3 |
| Posterior gills | 6 | < 5:30 |
|  | 12 | < 3 |
| Female + male gonads | 6 | < 8 |
|  | 12 | < 3 |
| Eye stalk + muscle | 6 | < 26 |
|  | 12 | < 19 |
| 1st + 2nd Zoea stage | 6 | < 11 |
|  | 12 | < 2 |
| 3rd Zoea stage | 6 | < 12 |
|  | 12 | < 6 |

Bowtie[19] aligner (version 0.12.9) was used to map back the raw short paired reads to the assembled transcripts produced by Trinity. The command-line parameters used with the *alignReads* script are shown in Dialogue 2. Table 4 provides a brief description of each parameter.

```
---left    "compatible_path_extension_for_reverse_reads"
--right    "compatible_path_extension_for _forward_reads"
--seqType fq --SS_lib_type RF --aligner bowtie –retain_intermediate_files  –target "compatible_path_extension_for trinity.fasta file" –Output "compatible_path_extension_for _output_folder".
```

*Dialogue 2. Command line arguments used with alignReads.pl script.*

*Table 4. Arguments used with alignReads and a brief description of each.*

| Argument | Brief Description |
| --- | --- |
| --aligner | The choice of aligner used; either bowtie or bowtie2 |
| -target | Path to the desired Trinity assembled transcriptome Fasta file |

Next, the Integrated Genomic Viewer (IGV)[20][21] was used to visualize the *Bam* alignment files generated by Bowtie and obtain assembly statistics for the raw reads that were able to be mapped back to one or more of the assembled transcriptomes in each tissue.

RSEM[22](RNA-Seq by Expectation –Maximization) v1.2.20 was used to estimate the gene and isoform expression levels in the assembled transcriptome files generated by Trinity.

## 2.3 Database and web server software

The original parsing of raw data was done using the Perl scripting language (version 5.12.4). To house our data, we used the MySQL (version 5.1.68) data-warehousing tool.

The project website utilizes a combination of PHP scripting (version 5.3.10), JavaScript (version 1.7.1), and HTML5 and it is powered by Kubuntu 14.04 operating system.

## 2.4 Blast database of sequences

We created twelve DNA Blast[23] databases for each of the raw (short reads) files that were produced by NGS sequencing. Furthermore, we created twelve DNA Blast databases for the assembled transcriptome files generated by Trinity ( note: two runs were performed for each one of the six tissue, each one with a different library type FR/RF). The DNA Blast databases are created using the command-line *makeblastdb* available as part of the blast+[24] package , with the command line parameters shown in Dialogue 3. Table 5 provides a brief description of each parameter.

```
-in "compatible_path_extension_for_input_file", -dbtype nucl, -out "compatible_path_extension_for_blast_database"
```

*Dialogue 3. Command line arguments used with makeblastdb.*

*Table 5. Arguments used with Trinity and a brief description of each.*

| Argument | Brief Description |
| --- | --- |
| -in | Input file name, the file in fasta format |
| -dbtype | Type of database; nucl for DNA database |
| -out | Name of blast database created |

## 2.5 Evaluation methods

With the absence of a reference genome, as in the case with non-model organism such as Uca minax, the process of evaluating the correctness and quality of the assembled transcriptomes via de novo sequence assembly methods becomes somewhat ambiguous. There is no definite criteria that can clearly draw a line that separates correct vs incorrect assembled transcriptomes. Hence, the decision was made to use house keeping genes as the criteria to judge the correctness of the assembled transcriptome. We hypothesize that if Trinity's assembly is indeed accurate then a blast search of the transcript database using these house keeping genes should yield alignments across all reconstructed tissues.

The house keeping genes of choice were Histone H3 and Ribosomal protein S16. Histone H3 is one of the five Histone proteins in eukaryotic cells. These proteins are the main components of chromatin which are responsible for packaging and ordering DNA into nucleosomes, as well as having a role in gene regulation. Histone proteins are among the most highly conserved proteins in eukaryotes. Ribosomal protein S16 as its name indicates is one of the proteins that, along with rRNA is responsible for building ribosomal units. S16 is the main protein used for reconstructing phylogenies due to it being highly conserved between different species.

For validation purposes, we used partial sequences of Uca minax Histone H3 and Ribosomal protein S16. These sequences are shown in Table 6. The partial sequences were cloned through the use of RT-PCR with degenerated primers (oligoas). However, considering the absence of a reference genome as well as the fact that the N-term and C-term of the corresponding proteins are rarely conserved, only partial sequences of the corresponding proteins could be retrieved from the Uca minax. Hence, two complete homologous sequences of evolutionary related organisms such as Lice (Pediculus humanus corporis) Histone H3 and Water flea Ribosomal S16 (Daphnia pules S16) were used to retrieve corresponding genes from the Uca minax transcriptome database generated by Trinity.

*Table 6. Sequence of Histone H3 and Ribosomal protein S16 genes.*

| Gene name | Sequence |
|---|---|
| Histone H3 | ATCTGCTCTGCTACCGGAGGAGTCAAGAAGCC CCACCGTTACAGGCCAGGCATCGCCGCACTGC GTGAAATCCGCCGCTACCAGAAGAGCACCGAG CTGCTCATCAGGAAGCTGCCTTTCCAGCGTCT GGTGCGCGAGATCGCCCAGGATTTCAAGACCG ATCTCCGCTTCCAGTCCTCTGCTGTCATGGCT CTCCAGGAGGCCTCAGAGGCTTACCTCGTCGG TCTCTTCGAGGACACCAACCTGTGCGATTTCC ACGCCATAGGGGGGGGAGTATAATAAAAGAGT GGGGTACGTTCACGCCGATTTAAGAAGATAGT GCAAAACGACTGCATAGGTATCCTGCTGTTTG AAGATCACACTCCAGTCTGTTACGCCACTCTT TATAAGACTAGTGGTTTTTGGGCCCGGCA |
| Ribosomal S16 subunit | TTGAGCCCAGGACACTGCAGTTCAAGTT GATGGAGCCTGTGTCGCTGCTGGGCAAG GAGAGGTTTTCCAATGTGTCCATCCGTG TGCGTGTGAAGGGTGGCGGACACACCTC CCAGGTCTATGCCATCCGTCAGGCCATC TCCAAGTCCCTCGTGGCTTACTACCAGA AGTTTGTGGACGAGGCCTCCAAGAAGG AGATCAAGAACATCCTTATCAACTATGA CAGGTCACTCTTGGTCGCTGACCCCAGG CGGTGTGAGCCCAAGAAGTTCGGAGGTC CTGGAGCCAGGGCACGCTACCAGAA |

## 3 Results and Discussion
### 3.1 Assembled transcriptome

In total, we created twelve Uca minax transcriptome libraries (six pairs). Each pair corresponds to the same tissue but with different assembly orientation (FR vs RF). As seen in Table 7, the number of transcripts assembled by Trinity for the same tissue slightly differs depending on the orientation. However, we noticed that for almost 80% of the transcripts Trinity assembles a sequence in one orientation (for example FR) and then assembles its reverse complement in the other orientation (RF). This signifies that the distinction of the direction of the raw reads might prove insignificant in our case. Moreover, Table 7 provides the basic statistics on the number of genes, isoforms and contigs assembled by Trinity for each run.

Further analysis of the assembled transcriptomes was needed to determine which transcripts were isoforms of the same gene, for that we used RSEM software. The results of the RSEM analysis are shown in Tables 8 and 9. The focus of the analysis was on two relative measure of transcript abundance: the Transcripts per million (TPM) and the Fragments per kilobase of exon per million reads mapped (FPKM) values. TPM indicates the number of transcripts of a specific type found if one million full transcripts from the sample are sequenced, given the abundances of the other transcripts in the sample. FPKM is the expected number of fragments to be found for each thousand bases in the feature for every $N/10^6$ sequenced fragment if the same RNA pool was sequenced again. The results of the gene quantification analysis for the Uca minax transcriptomes are shown in Table 8. This table enumerates the number of genes as well as their percentage in each library of assembled transcriptome that have FPKM, TPM and expected count values equal zero. On the other hand, Table 9 contains the number of assembled transcripts as well as their percentage in each library of the assembled transcriptome that have FPKM, TPM, expected count and IsoPCT (which stands for the percentage of this transcript's abundance over its parent gene's abundance) values of zero.

*Table 7. Basic Trinity statistics*

| Library name | Read Orientation | Total Trinity Transcripts | Trinity Components | Contig N50 |
|---|---|---|---|---|
| Anterior gills | FR | 108674 | 87591 | 779 |
| | RF | 108651 | 87569 | 770 |
| Posterior gills | FR | 117370 | 90888 | 876 |
| | RF | 117173 | 90832 | 876 |
| Female + male gonads | FR | 118288 | 93389 | 708 |
| | RF | 118397 | 93498 | 695 |
| Eye stalk + muscle | FR | 169817 | 156547 | 292 |
| | RF | 168591 | 155342 | 291 |
| 1st + 2nd Zoea stage | FR | 152489 | 120625 | 717 |
| | RF | 152197 | 120744 | 716 |
| 3rd Zoea stage | FR | 165495 | 119072 | 1088 |
| | RF | 165895 | 119081 | 1081 |

*Table 8. Uca minax RSEM analysis (genes)*

| Library name | Read orientation | FPKM =0 |
|---|---|---|
| Anterior gills | FR | 22301 (26%) |
| | RF | 1424 (2%) |
| Posterior gills | FR | 21394 (24%) |
| | RF | 21739 (25%) |
| Female + male gonads | FR | 33608 (38%) |
| | RF | 33509 (38%) |
| Eye stalk + muscle | FR | 58417 (67%) |
| | RF | 58254 (67%) |
| 1st + 2nd Zoea stage | FR | 31754 (36%) |
| | RF | 31885 (36%) |
| 3rd Zoea stage | FR | 24802 (28%) |
| | RF | 24979 (28%) |

*Table 9. Uca minax RSEM analysis (isoforms)*

| Library name | Read orientation | FPKM =0 |
|---|---|---|
| Anterior gills | FR | 30581 (28%) |
|  | RF | 7345 (7%) |
| Posterior gills | FR | 29896 (28%) |
|  | RF | 30120 (28%) |
| Female + male gonads | FR | 45051 (41%) |
|  | RF | 44795 (41%) |
| Eye stalk + muscle | FR | 74543 (67%) |
|  | RF | 74339 (67%) |
| 1st + 2nd Zoea stage | FR | 42744 (40%) |
|  | RF | 42881 (39%) |
| 3rd Zoea stage | FR | 34862 (32%) |
|  | RF | 35024 (32%) |

## 3.2  Public web resources

All resources and tools are publicly available via our website (www.rdc.cse.sc.edu/Uca_minax/TransNav ). This section will highlight the functionality of each tab accessible from the main menu.

*Blast*—We provide an easy-to-use interface for Blast search conducted on a comprehensive set of databases. A Blast search can be performed on both the raw data and the reconstructed transcriptome data obtained via Trinity. The search tool allows for the user to either upload a file containing their query sequences (in FASTA format) or simply copy and paste their sequences into the provided text box. To perform searches on multiple data sets at once, the user can simply check the boxes beside each desired database. Once the submit button is clicked our engine will perform a Blast search on each database sequentially and display the alignment results for each database below the search tool in separate windows. The user can then choose to download a txt or html version of the results to save for future reference, see Figure 1.

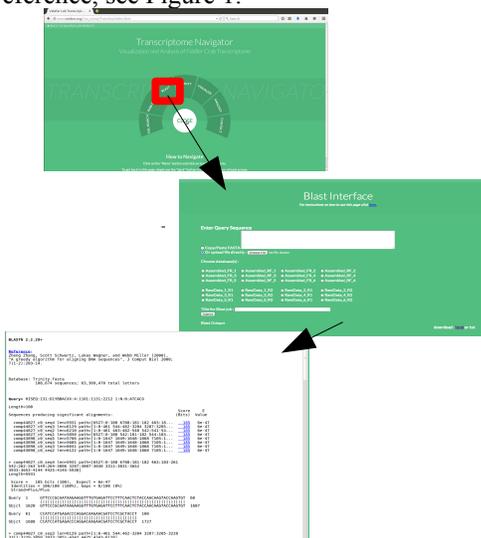

Figure 1: Blast Query result example

*Visualize*—To evaluate the quality of the reconstructed transcripts and assess results of the Trinity software package, we use Bowtie to align the resulting transcriptome to the original 100 bp reads. IGV (Integrated Genome Viewer)[5][6] is used to view the subsequent alignments. In doing this, we can get an idea of how well Trinity has conserved the original reads in its reconstruction as well as establish a confidence in the reconstructed transcriptome based on the observed coverage. The user can select any of the links on the "Visualization Tool" screen to view the alignment. Doing so will open IGV on their local machine and automatically load the Trinity results along with its alignment to the original reads, see Figure 2. The IGV package can be downloaded from the Broad Institute's website at http://broadinstitute.org/igv/home.

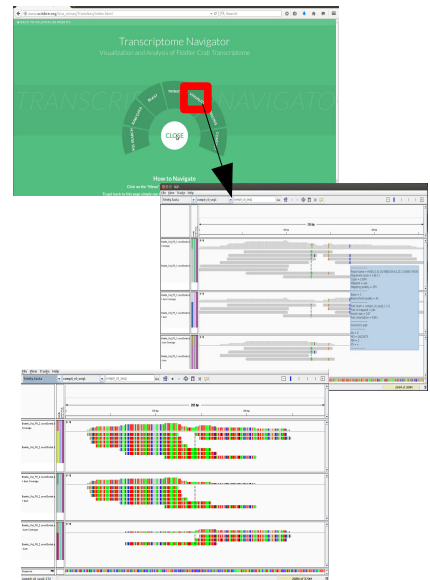

Figure 2: IGV alignment example

*Trinity*—We provide links for direct download of all Trinity results. On the "Trinity" screen each tissue has two download options. Clicking on the "FASTA" link will download the Trinity assembled transcriptome library for that tissue in FASTA format. Selecting the "Blast searchable database" option will download the data sets in a convenient Blast searchable format. After clicking the links, the files will be downloaded and saved to the user's local machine.

*Analysis*—On the "Analysis" page of the website the user can find gene and isoform abundance reports generated from the Trinity results for each individual tissue. These files are available for download and are in a spreadsheet-friendly, tab delimited format. The abundance reports were generated using the software package RSEM[7] .

*Raw Data*—We have also made all of our raw sequenced data available for download. The raw data sets can be downloaded in three forms: FASTA, FASTQ and Blast searchable databases. Each format can be accessed by selecting their respective links under each tissue.

*SQL Search*—The "SQL Search" tab leads to a page that directs the user to the phpMyAdmin

([http://www.phpmyadmin.net/](http://www.phpmyadmin.net/)) view of our databases. Select collaborators have been granted access to perform their own SQL queries on our data. Both the original database and compressed database are available to explore.

*Contact*—All contact information for all collaborators and labs are available on this page.

## 3.3 Validation of house keeping genes

Blastx[23] was used to find a matching transcriptome for both the partial sequences and the complete ones. In the Histone H3 case, only two transcriptome libraries produced a match. As shown in Table 10, the two matching transcriptomes were found in both the female and male gonads and the eye stalk and muscle libraries. All matching transcriptomes from the female and male gonad libraries (in both read directions) matched both the complete and partial sequence of the Histone H3. On the other hand, the results from the Eye stalk and muscle libraries varied from one read direction to the other. In the forward/reverse (FR) direction, four transcriptomes were matched but only three of them matched both the complete and partial Histone H3 sequences. The last transcriptome matched only with the partial Histone H3 sequence. In the reverse/forward (RF) direction, three matches were found. Only two matched both the complete and partial H3 sequences while the last transcriptome matched the partial H3 sequence.

In the case of Ribosomal S16 protein, two matches were found in each one of the six libraries for the partial S16 sequence, however none of the assembled transcriptomes in all six libraries matched the complete Water flea sequence, for more details see Table 11. In both Table 10 and 11;the first column identifies the transcriptome library name, the second column specifies the assembly orientation, the third column lists the transcriptome id assigned by Trinity during the assembly process, the fourth column contains the Bp length of the assembled transcriptome, the fifth column is the scoring assigned by Blast, the sixth column is the percentage of matching identities between the transcript and the query sequence.

*Table 10: Transcriptome match results for the Histone H3, where RD is the Read Direction and Trans. ID is the Transcriptome ID*

| Library | RD | Trans. ID | Len. | Score | Identities |
|---|---|---|---|---|---|
| Female + male Gonads | FR | comp106449_c0_seq1 | 227 | (Complete) 71 | 129/158 (82%) |
| | | | | (Uca) 82 | 83/84 (99%) |
| | | comp144323_c0_seq1 | 398 | (Complete) 56 | 92/110 (84%) |
| | | | | (Uca) 81 | 83/84 (99%) |
| | RF | comp107689_c0_seq1 | 227 | (Complete) 71 | 129/158 (82%) |
| | | | | (Uca) 116 | 118/119 (99%) |
| | | comp133719_c0_seq1 | 398 | (Complete) 56 | 92/110 (84%) |
| | | | | (Uca) 81 | 83/84 (99%) |
| Eye stalk + muscle | FR | comp107816_c0_seq1 | 384 | (Complete) 86 | 172/215 (80%) |
| | | | | (Uca) 166 | 168/169 (99%) |
| | | comp111814_c1_seq2 | 448 | (Complete) 77 | 146/180 (81%) |
| | | | | (Uca) 141 | 143/144 (99%) |
| | | comp111814_c1_seq1 | 512 | (Complete) 77 | 146/180 (81%) |
| | | | | (Uca) 141 | 143/144 (99%) |
| | | comp68227_c0_seq1 | 257 | (Uca) 109 | 118/122 (97%) |
| | RF | comp106949_c0_seq1 | 484 | (Complete) 86 | 172/215 (80%) |
| | | | | (Uca) 166 | 168/169 (99%) |
| | | comp112248_c1_seq1 | 472 | (Complete) 77 | 146/180 (81%) |
| | | | | (Uca) 141 | 143/144 (99%) |
| | | comp71139_c0_seq1 | 258 | (Uca) 109 | 118/122 (97%) |

## 3.4 Analysis of data quality

As a result of the house keeping gene validation step, it was important to investigate the quality of the original raw sequenced data to produce a suitable per-processing template to further improve results of the de novo assembly step. The quality control analysis of the raw sequenced data was performed using FastQC[25] (v0.11.2) software package. All raw sequenced data files used Sanger/Illumina 1.9 encoding. Table 12 contains some of the basic statistics provided by FastQC analysis of the original raw sequenced data. The percentage of nitrogenous bases on a DNA/RNA molecule that are either guanine or cytosine is known as GC content. Determination of this ratio contributes in mapping gene-rich regions of the genome. Overall GC content of all the bases of all sequences and total sequences are two of the parameters that we report in this table. Both of these parameters are the same for the forward read and the reverse reads of the same tissue.

*Table 11: Transcriptome match results for the Ribosomal S16*

| Library | RD | Trans. ID | Len. | Score | Identities |
|---|---|---|---|---|---|
| Anterior gills | FR | comp22755_c0_seq1 | 548 | 81 | 301/304 (99%) |
| | | comp22753_c0_seq1 | 544 | 295 | 301/304 (99%) |
| | RF | comp38036_c1_seq1 | 547 | 295 | 301/304 (99%) |
| | | comp38036_c0_seq1 | 521 | 295 | 301/304 (99%) |
| Posterior gills | FR | comp25671_c0_seq1 | 548 | 295 | 301/304 (99%) |
| | | comp25635_c0_seq1 | 550 | 295 | 301/304 (99%) |
| | RF | comp42515_c1_seq1 | 548 | 295 | 301/304 (99%) |
| | | comp42515_c0_seq1 | 523 | 295 | 301/304 (99%) |
| Female + male gonads | FR | comp41201_c0_seq1 | 545 | 298 | 301/304 (99%) |
| | | comp45019_c0_seq1 | 545 | 295 | 301/304 (99%) |
| | RF | comp49683_c0_seq1 | 522 | 298 | 302/304 (99%) |
| | | comp49683_c1_seq1 | 544 | 295 | 301/304 (99%) |
| Eye stalk + muscle | FR | comp85313_c0_seq1 | 503 | 295 | 301/304 (99%) |
| | | comp81645_c0_seq1 | 459 | 295 | 301/304 (99%) |
| | RF | comp89360_c0_seq2 | 526 | 298 | 302/304 (99%) |
| | | comp89360_c0_seq1 | 461 | 298 | 302/304 (99%) |
| 1st + 2nd Zoea stage | FR | comp52715_c0_seq1 | 552 | 295 | 301/304 (99%) |
| | | comp68619_c1_seq1 | 522 | 292 | 300/304 (99%) |
| | RF | comp63400_c0_seq1 | 519 | 295 | 301/304 (99%) |
| | | comp63400_c1_seq1 | 528 | 292 | 300/304 (99%) |
| 3rd Zoea stage | FR | comp67340_c1_seq1 | 542 | 295 | 301/304 (99%) |
| | | comp67340_c0_seq1 | 513 | 295 | 301/304 (99%) |
| | RF | comp54913_c0_seq1 | 545 | 295 | 301/304 (99%) |
| | | comp54911_c0_seq1 | 543 | 295 | 301/304 (99%) |

*Table 12: FastQC Basic Statistics*

| Tissue | GC% | Total sequences |
|---|---|---|
| Anterior gills | 46 | 7360115 |
| Posterior gills | 44 | 10870260 |
| Female + male gonads | 41 | 8308628 |
| Eye stalk + muscle | 39 | 9761180 |
| 1st + 2nd Zoea stage | 43 | 8481444 |
| 3rd Zoea stage | 47 | 9973433 |

Analysis for the original raw sequenced data revealed the same patterns for five of the six tissues sequenced (anterior gills, posterior gills,, female+male gonads, 1st + 2nd zoea stage and 3rd zoea stage). These patterns are listed below.

1. Good scoring results for the following modules: per base sequence quality, per sequence quality score, per base N content, sequence length distribution and adapter content.

2. Warnings were issued for sequence duplication levels, overrepresented sequences modules and some files issued warnings for the per tile sequence quality module.

3. Failed scoring results for the following modules: per base sequence content, per sequence GC content and kmer content.

On the other hand, the eye stalk + muscle tissue deviated from this pattern by having good scoring results for four modules only: per sequence quality score, per base N content and sequence length distribution Warnings were issued for sequence duplication levels, per sequence GC content modules and the R1 file issued warnings for the per tile sequence quality module. All of the remaining modules failed the scoring result.

For the analysis of the five tissues with the same pattern, it showed that the libraries were biased in the first 15bp of the raw sequences and that some contamination of the libraries occurred mainly because of adapter sequences and PCR primer sequences which are normally part of next generation sequencing protocol, existed as part of the raw sequences. This can be easily solved by trimming both the adapter and PCR primer sequences and removing any primer-dimer in the raw libraries, also some sequence truncation may be performed later on. However, for the eye stalk and muscle tissue case, it seems that the adapter contamination is more severe and although it can be lessened by trimming, the resulting raw sequences after trimming may not provide enough coverage for a good transcripts assembly result.

# 4 Conclusion

Our initial investigation of the Uca minax transcriptome assembly indicates successful reconstruction for five of the six tissue with some potential room for improvement. The five well behaved tissues include: anterior gills, posterior gills,, female+male gonads, 1st + 2nd zoea stage and 3rd zoea stage. Data from these tissue samples can be further improved for a second round of assembly by removal of the primer and adapter sequences. This process will require a preprocessing step where a-priori knowledge of these sequences is used to remove redundant primer/adapter sequences. Upon cleansing of the raw data, we speculate that transcriptome assembly will result in better reconstructed genetic data. The same process will be repeated for the "Eye stalk + muscle" that exhibits the most degree of contamination. We speculate that preprocessing of the data will improve the quality of the final assembled transcriptome, but will likely contain less number and shorter genes. We will publish our final assembled genome at our current website.

# 5 Acknowledgements

This work was supported by NIH Grant Number P20 RR-016461 to Dr. Homayoun Valafar and INBRE Pilot Project grant to Dr. Giuliana Gusmaroli.